\documentclass[a4paper]{article}
\usepackage{graphicx}
\usepackage{subfig}
\usepackage{cite}

\title{Invariants of Spin Networks from Braided Ribbon Networks}
\author{Jonathan Hackett}

\begin{document}

\maketitle

\begin{abstract}
We connect Braided Ribbon Networks to the states of loop quantum gravity. Using this connection we present the reduced link as an invariant which captures information from the embedding of the spin-networks. We also present a means of understanding higher valent nodes in the context of braided ribbon networks and an interpretation of the dual of these nodes as polygons or polyhedra.
\end{abstract}

\section{Introduction}

Braided Ribbon Networks are an extension of the spin network states of loop quantum gravity originally intended as a framework in which particle physics could emerge.\cite{BilsonThompson:2006yc} The hope of having particle physics emerge from quantum gravity spawned a significant amount of work divided between two programs: three valent braided ribbon networks \cite{BilsonThompson:2006yc, Hackett:2007dx,Hackett:2008ie, BilsonThompson:2008ex,  BilsonThompson:2009fh} and four valent braided ribbon networks \cite{Wan:2007nf,Smolin:2007sn,He:2008is, He:2008jc, Wan:2008qs, Hackett:2008tt}. In \cite{Hackett2011} these two approaches were brought together into a single consistent framework.

Here we will take this unification further. Reminding the reader of the form of braided ribbon networks, we will expand this structure to higher valence braided ribbon networks. After this we will demonstrate a correspondence between these states and the states of loop quantum gravity. Using this correspondence we will then show that the reduced link as defined in \cite{Hackett2011} is in fact an invariant of the embedded spin network states of loop quantum gravity.

\section{Braided Ribbon Networks}
We follow the definition for Braided Ribbon Networks as laid out in \cite{Hackett2011}.
\begin{quote}
We begin by considering an $n$-valent graph embedded in a compact $3$ dimensional manifold. We construct a $2$-surface from this by replacing each node by a $2$-sphere with $n$ punctures ($1$-sphere boundaries on the $2$-sphere), and each edge by a tube which is then attached to each of the nodes that it connects to by connecting the tube to one of the punctures on the $2$-sphere corresponding to the node.

Lastly we add to each tube $n-1$ curves from one puncture to the other and then continue these curves across the sphere in such a way that each of the $n$ tubes connected to a node shares a curve with each of the other tubes.

We will freely call the tubes between spheres \textit{edges}, the spheres \textit{nodes} and the curves on the tubes \textit{racing stripes} or less formally \textit{stripes}.

We will call a Braided Ribbon Network the equivalence class of smooth deformations of such an embedding that do not involve intersections of the edges or the racing stripes.
\end{quote}

\begin{figure}[!h]
  \begin{center}
    \includegraphics[scale=0.2]{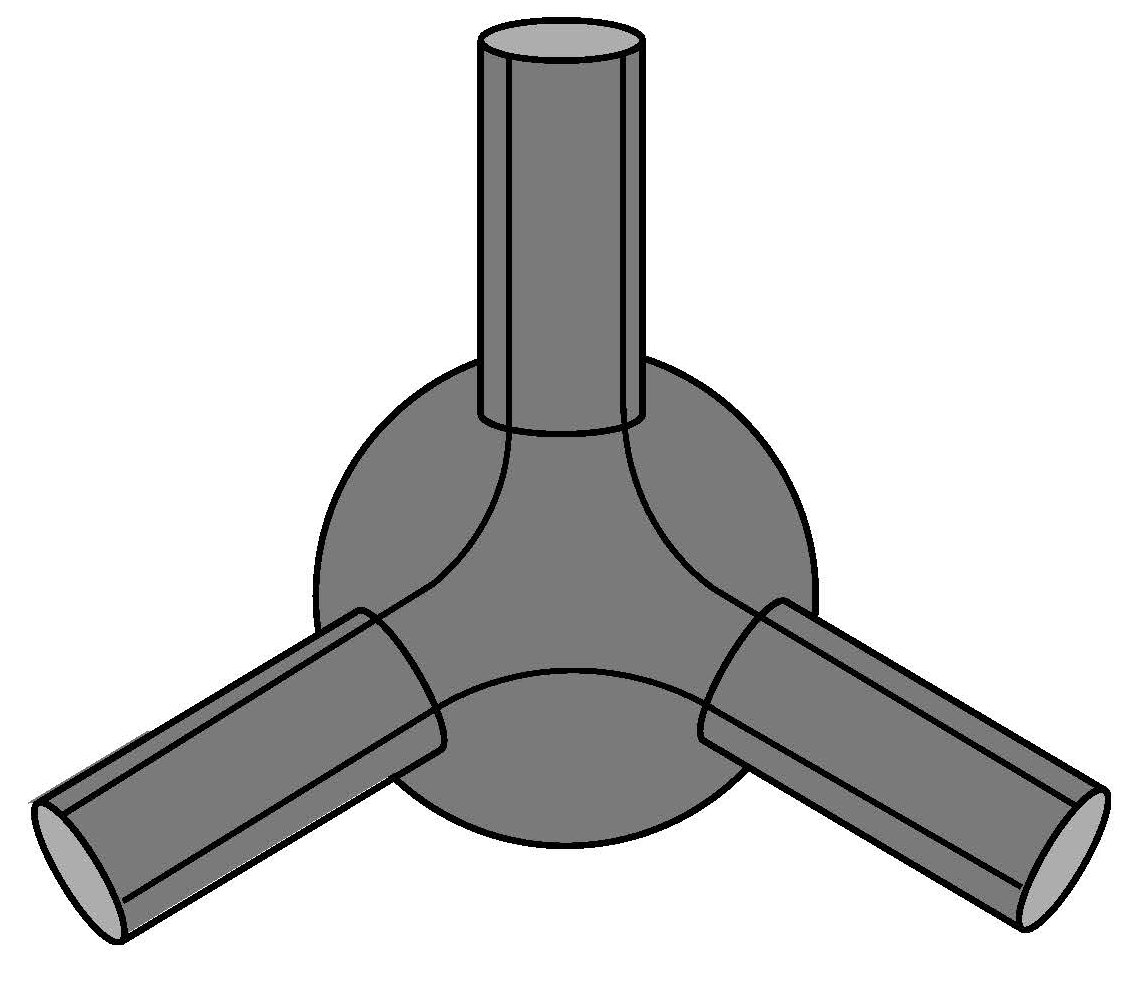}
  \end{center}
\caption{Three Valent Node}
\end{figure}

\begin{figure}[!h]
  \begin{center}
    \includegraphics[scale=0.2]{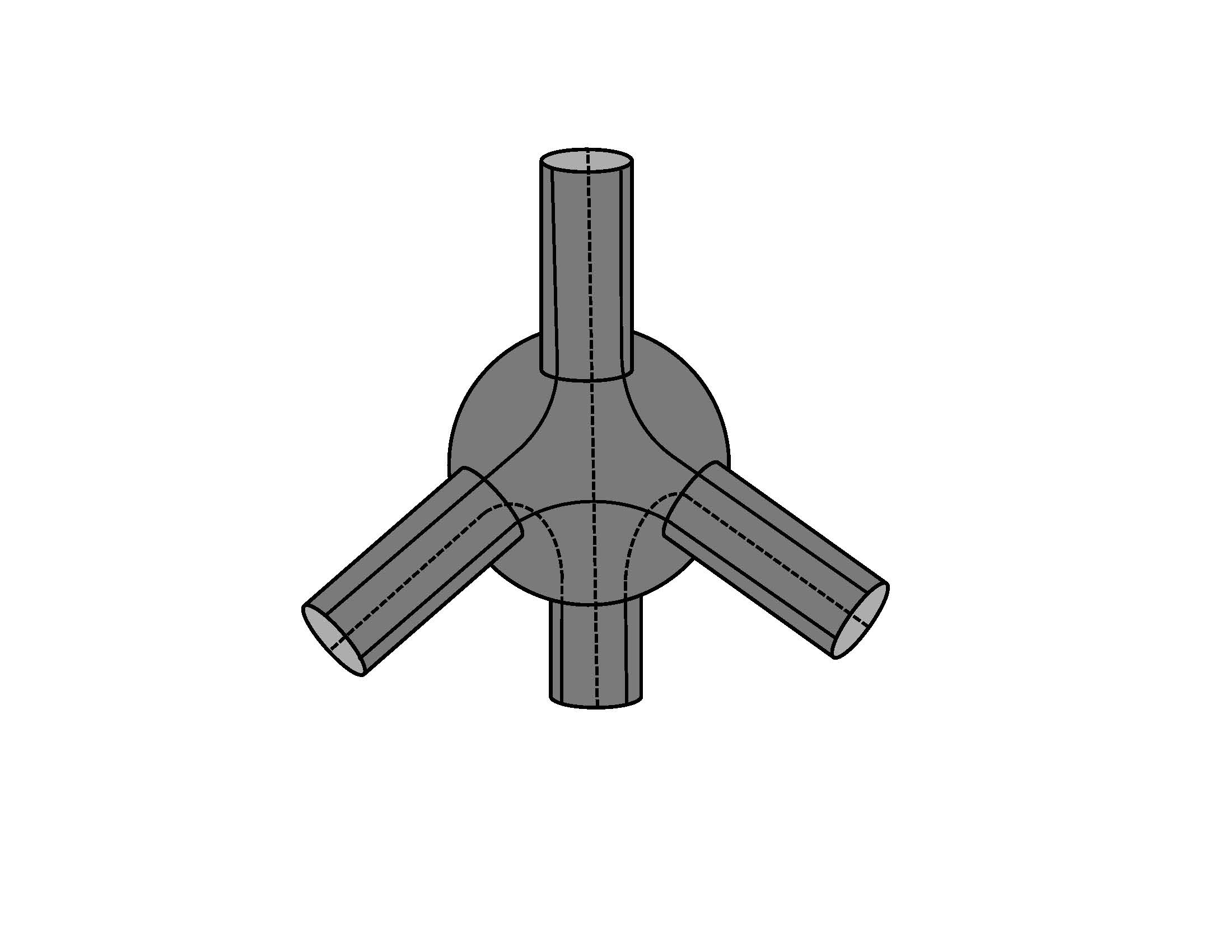}
  \end{center}
\caption{Four Valent Node}
\end{figure}
\subsection{The Evolution Moves}
With the duality of $n$ valent nodes to $n-1$ simplices established in \cite{Hackett2011} we define evolution moves on the graph by making reference to the Pachner moves on the gluings of the dual simplices. The operations are defined as preserving the external identifications of sub simplices corresponding to the racing stripes along the tubes, and we will go through these explicitly for both $3$ and $4$ valent BRNs.

\begin{figure}[!h]
  \begin{center}
    \includegraphics[scale=0.2]{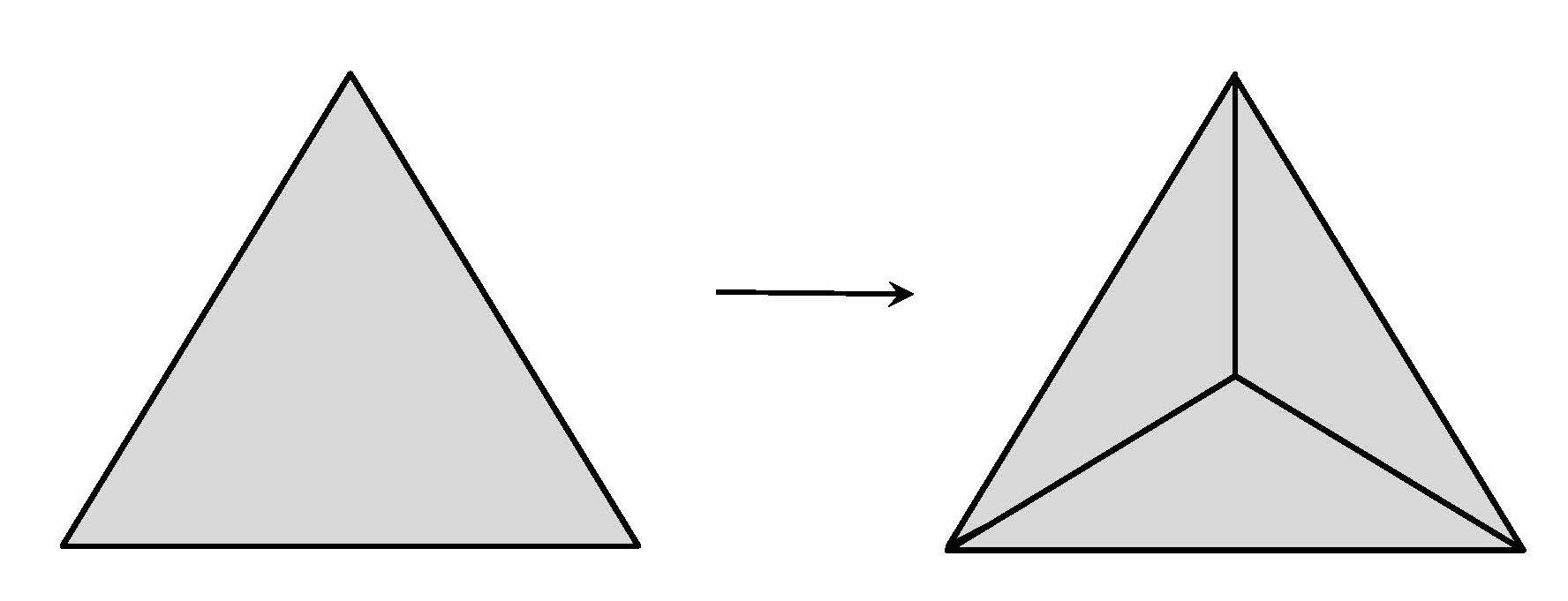}
  \end{center}
\caption{The 1-3 Pachner move} \label{4}
\end{figure}

\begin{figure}[!h]
  \begin{center}
    \includegraphics[scale=0.2]{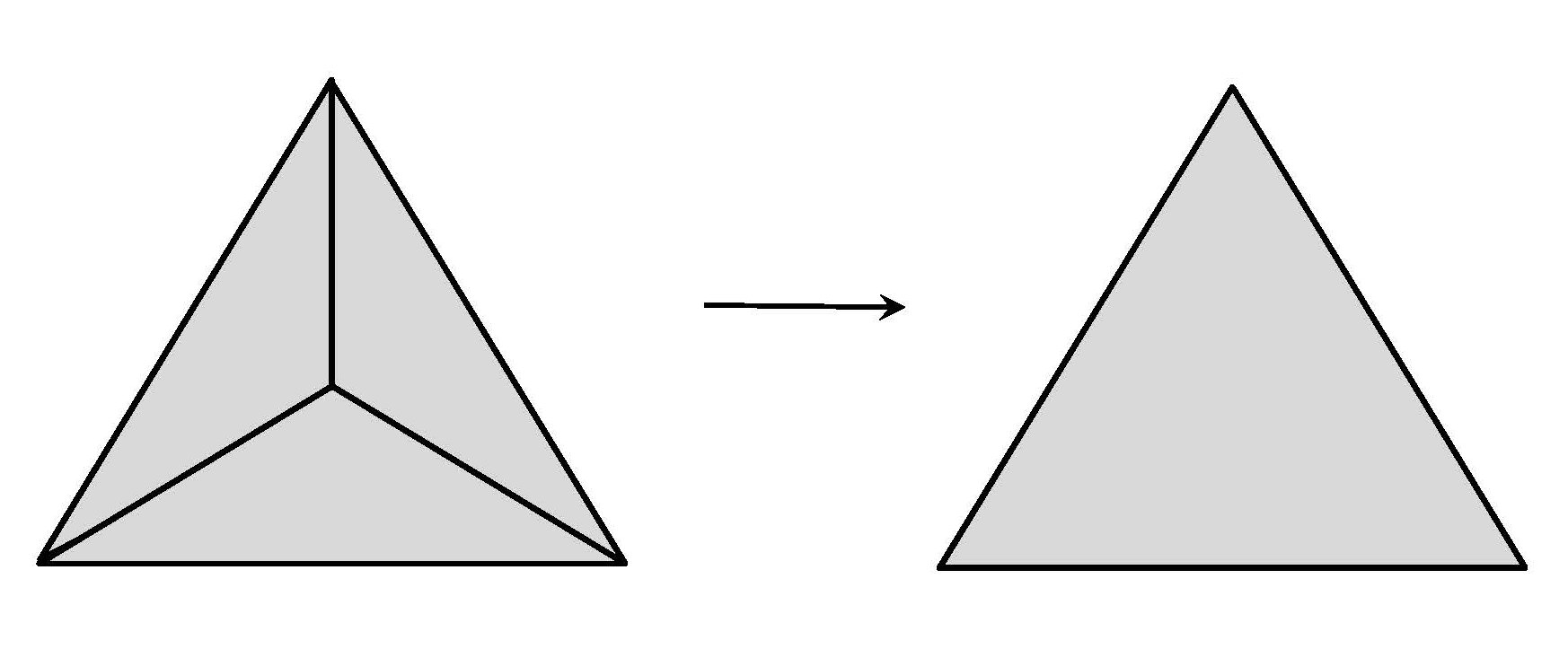}
  \end{center}
\caption{The 3-1 Pachner move} \label{5}
\end{figure}

\begin{figure}[!h]
  \begin{center}
    \includegraphics[scale=0.2]{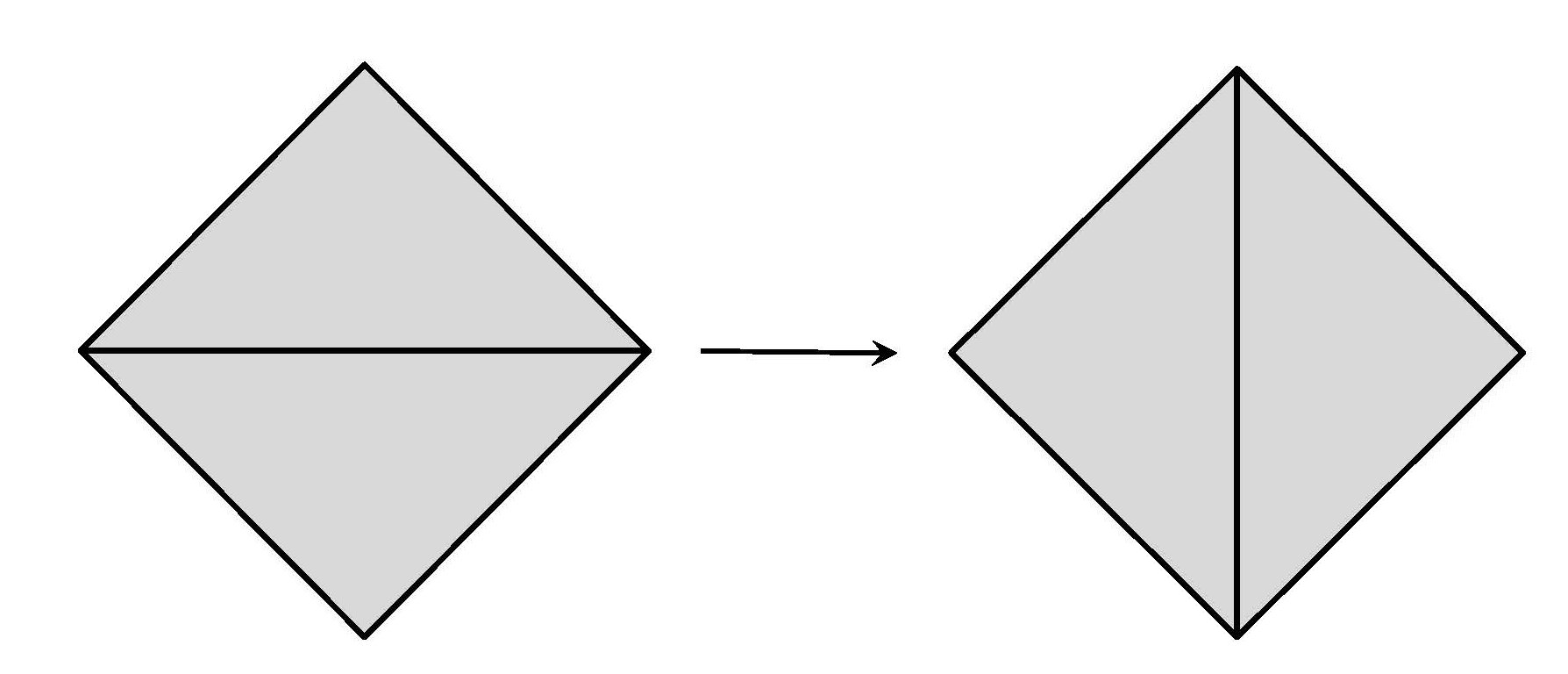}
  \end{center}
\caption{The 2-2 Pachner move} \label{6}
\end{figure}

\begin{figure}[!h]
  \begin{center}
    \includegraphics[scale=0.2]{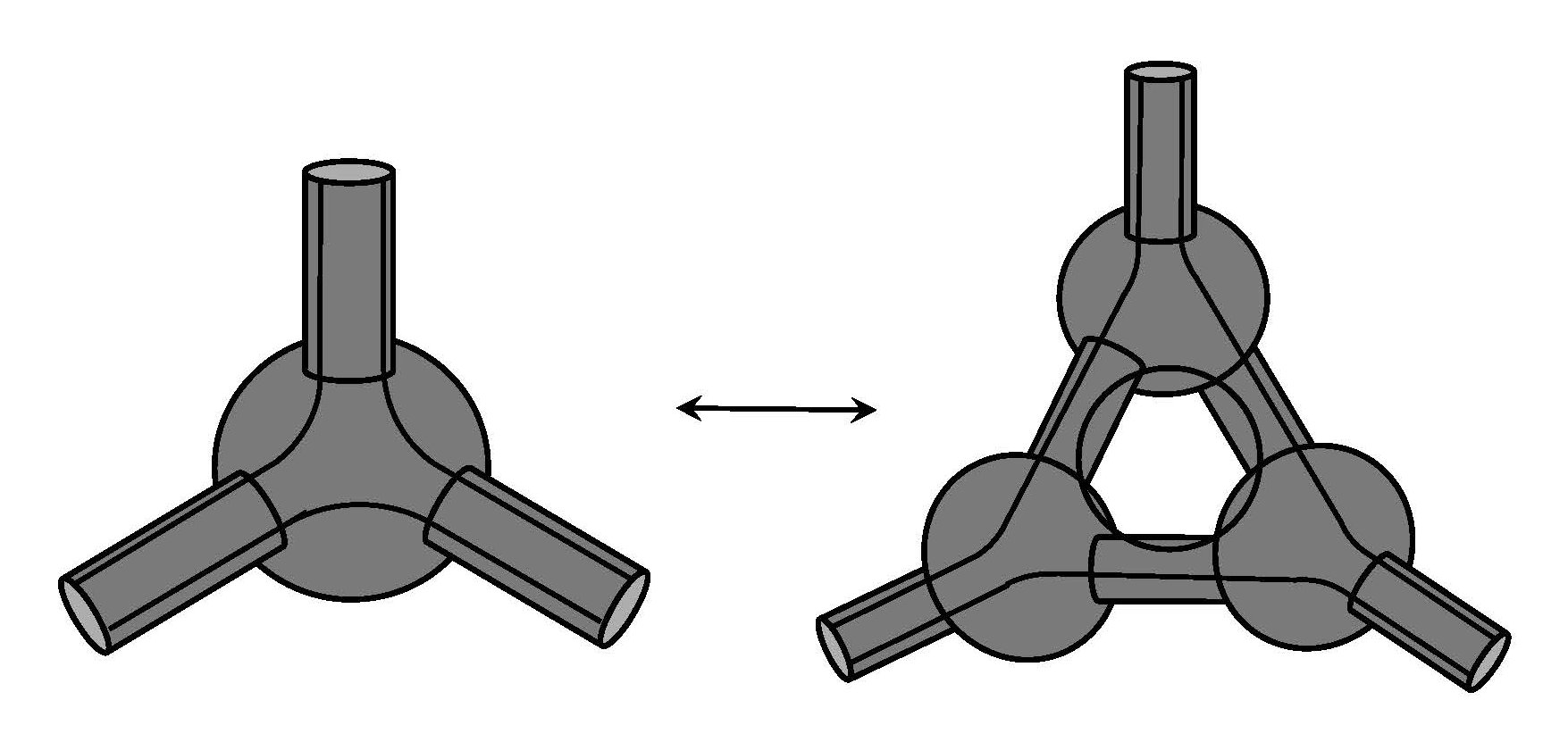}
  \end{center}
\caption{The 1-3 and 3-1 evolution moves} \label{7}
\end{figure}

\begin{figure}[!h]
  \begin{center}
    \includegraphics[scale=0.2]{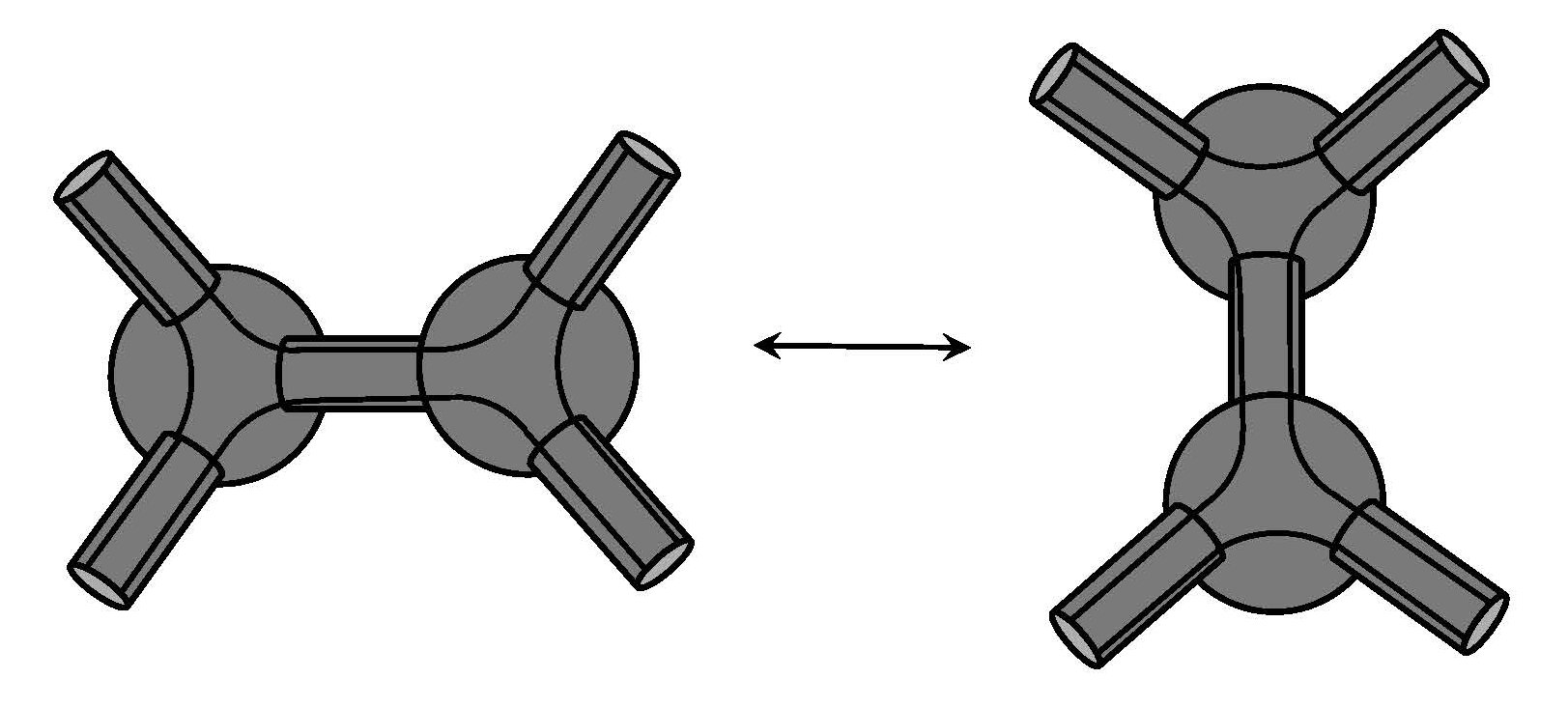}
  \end{center}
\caption{The 2-2 evolution move} \label{8}
\end{figure}

For $3$ valent networks, the corresponding simplices are triangles. The Pachner moves on triangles are the $1-3$ move (figure \ref{4}), the $3-1$ move (figure \ref{5}) and the $2-2$ move (figure \ref{6}). These correspond to similarly named evolution moves for the braided ribbon networks (figures \ref{7} and \ref{8}).

\begin{figure}[!h]
  \begin{center}
    \includegraphics[scale=0.2]{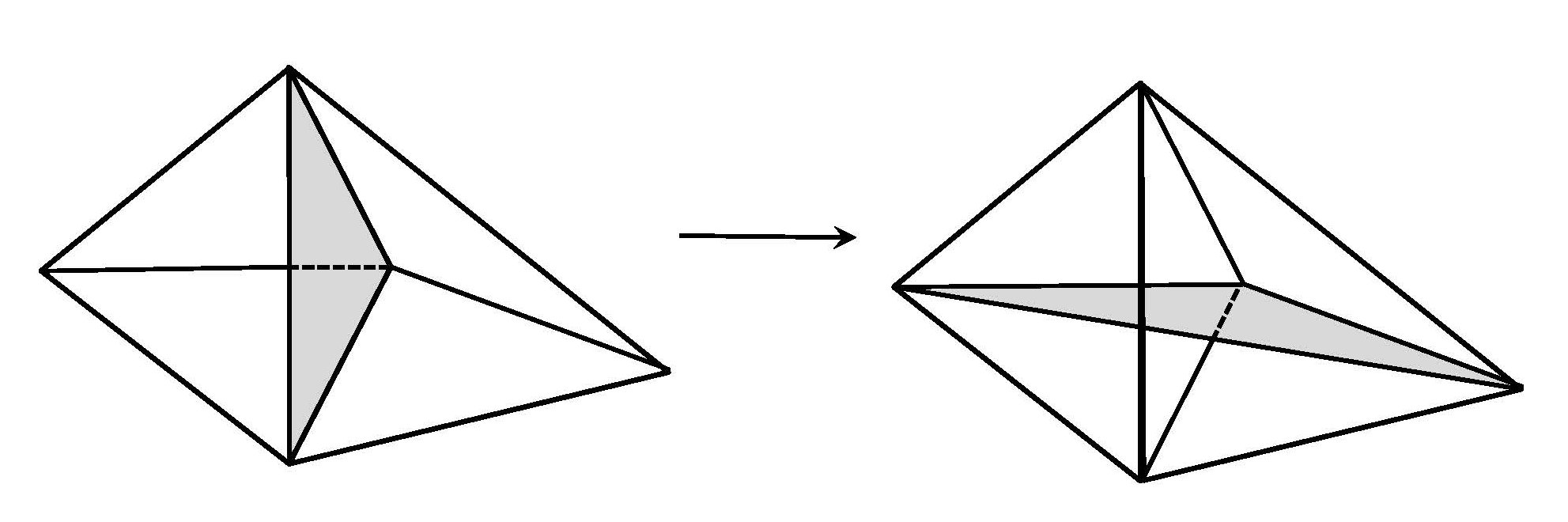}
  \end{center}
\caption{The 2-3 and 3-2 Pachner moves} \label{9}
\end{figure}

\begin{figure}[!h]
  \begin{center}
    \includegraphics[scale=0.2]{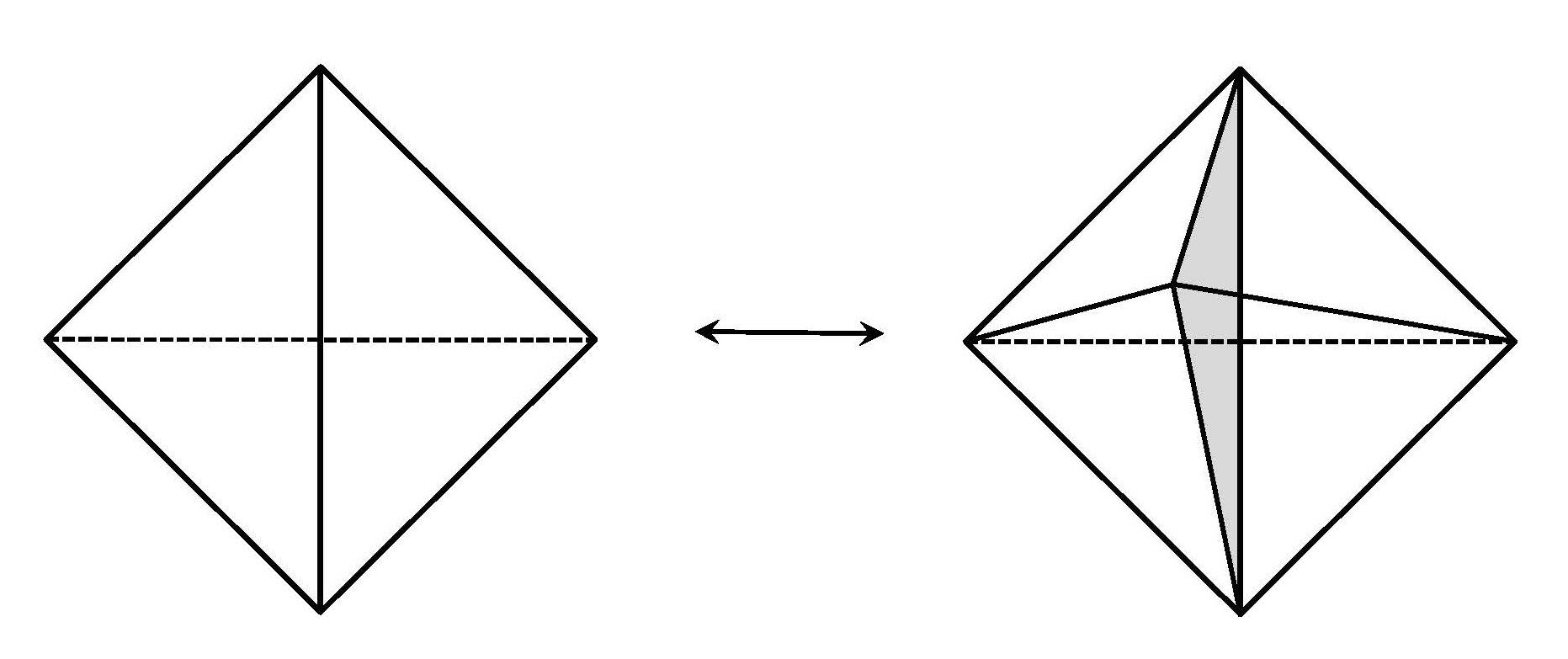}
  \end{center}
\caption{The 1-4 and 4-1 Pachner move} \label{10}
\end{figure}

\begin{figure}[!h]
  \begin{center}
    \includegraphics[scale=0.2]{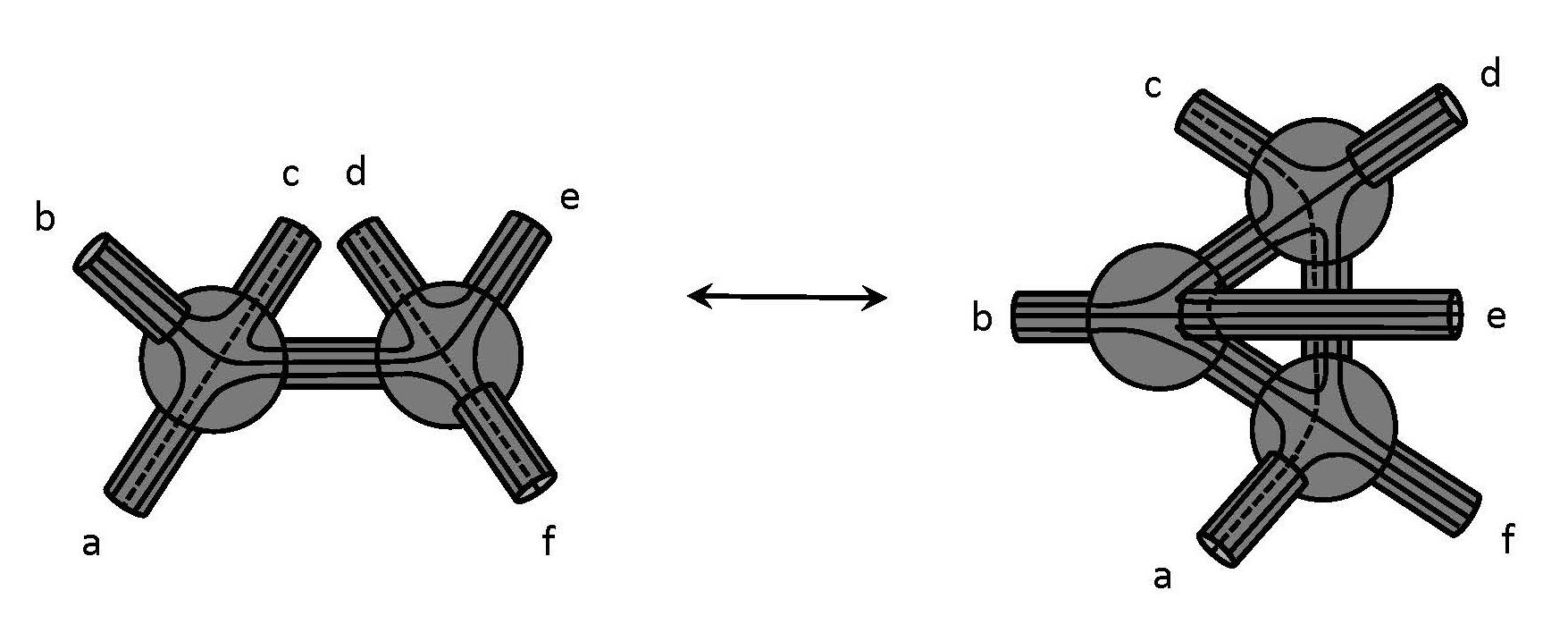}
  \end{center}
\caption{The 2-3 and 3-2 evolution moves} \label{11}
\end{figure}

\begin{figure}[!h]
  \begin{center}
    \includegraphics[scale=0.2]{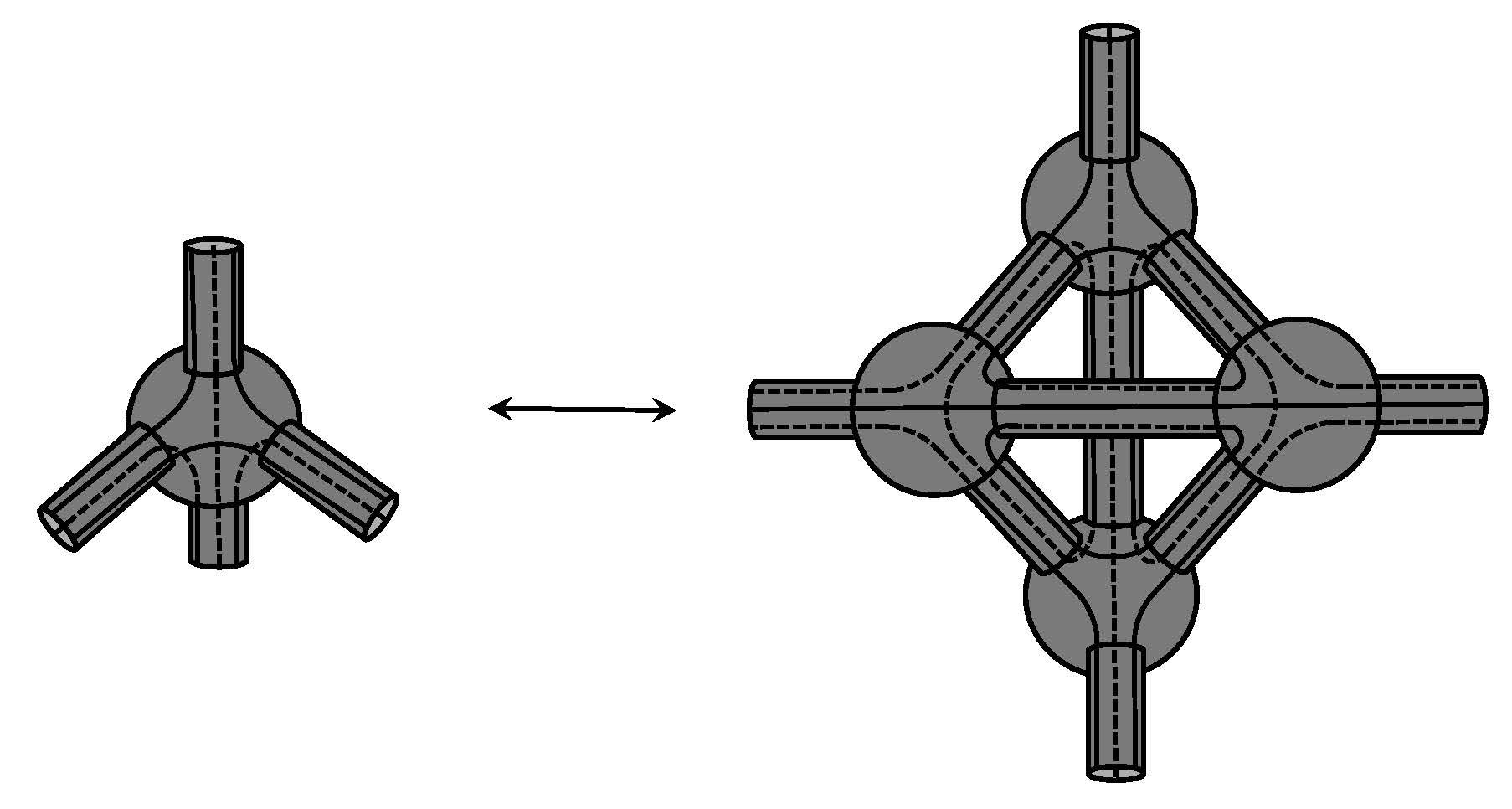}
  \end{center}
\caption{The 1-4 and 4-1 evolution moves} \label{12}
\end{figure}

For $4$ valent networks, the corresponding simplices are tetrahedra. Here the Pachner moves are the $2-3$ and $3-2$ moves (figure \ref{9}), and the $1-4$ and $4-1$ moves (figure \ref{10}). The corresponding evolution moves on braided ribbon networks are then figures \ref{11}, and \ref{12}.

In all of these cases we require that the moves are only allowed if they can be performed smoothly, that is to say that the operation can be performed through smooth deformations of the 2-surface of the BRN together with point-like changes in genus of the surface which occur at points which are in the interior of the $3$-manifold. This restricts us from performing something like the $1-3$ move in such a manner that the three nodes now encircle some other part of the 2-surface or a feature of the topology of the 3-manifold (which may occur if the space were a 3 dimensional torus for example).

\subsection{Higher Valence BRNs}
In \cite{Hackett2011}, it was shown that there do not exist any BRNs of valence higher than $4$. This fact follows from the four-colour theorem (or equally well Kuratowski's theorem) and the correspondence between $4$ valent vertices and tetrahedra. We now will introduce a modification to the framework that allows for higher valence vertices.

To do this we first make a few definitions.

\begin{quote}
We define the \textit{natural valence} of a braided ribbon network to be the number of racing stripes on each edge.
\\
We say that a node is \textit{natural} if each of the tubes which intersect share a racing stripe with each of the other tubes.
\\
Otherwise we will say that a node is \textit{composite}.
\end{quote}

\begin{figure}[!h]
  \begin{center}
    \includegraphics[scale=0.2]{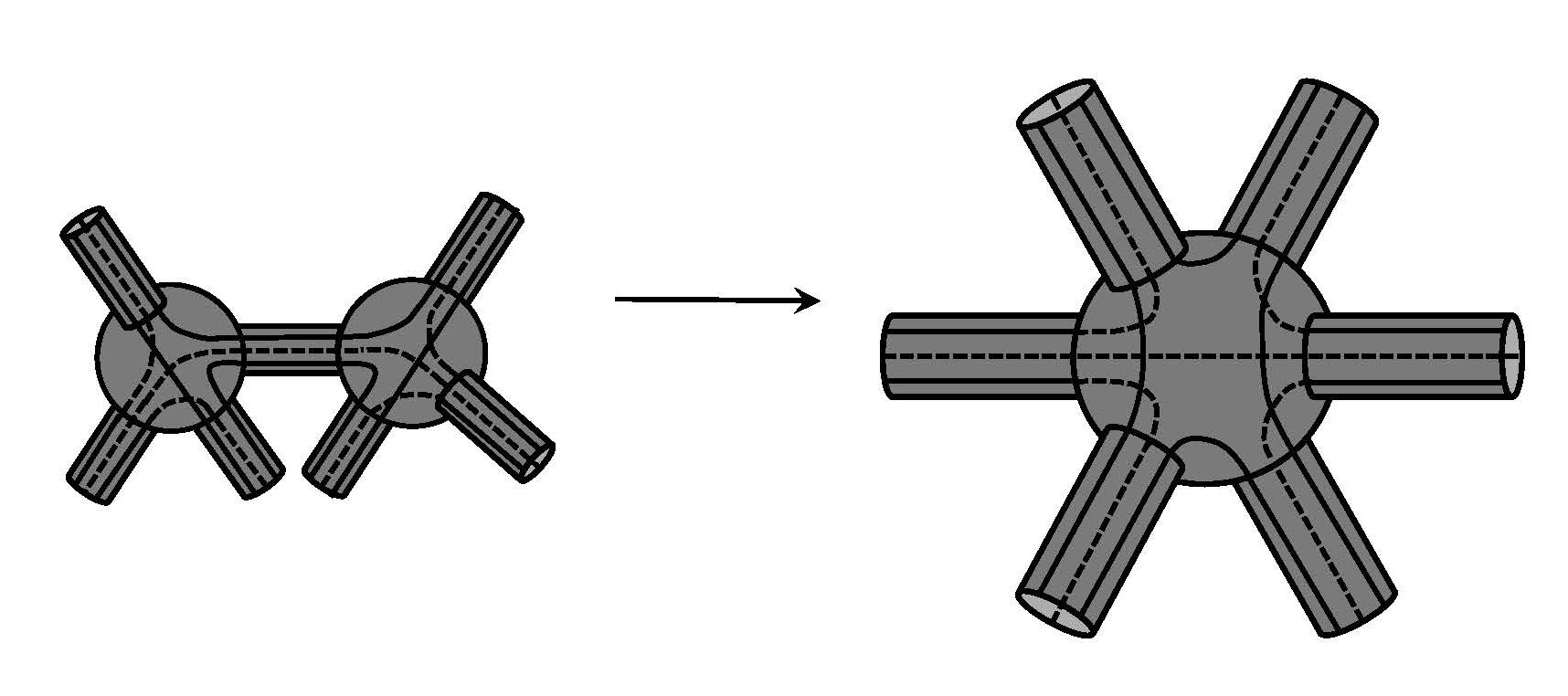}
    \includegraphics[scale=0.2]{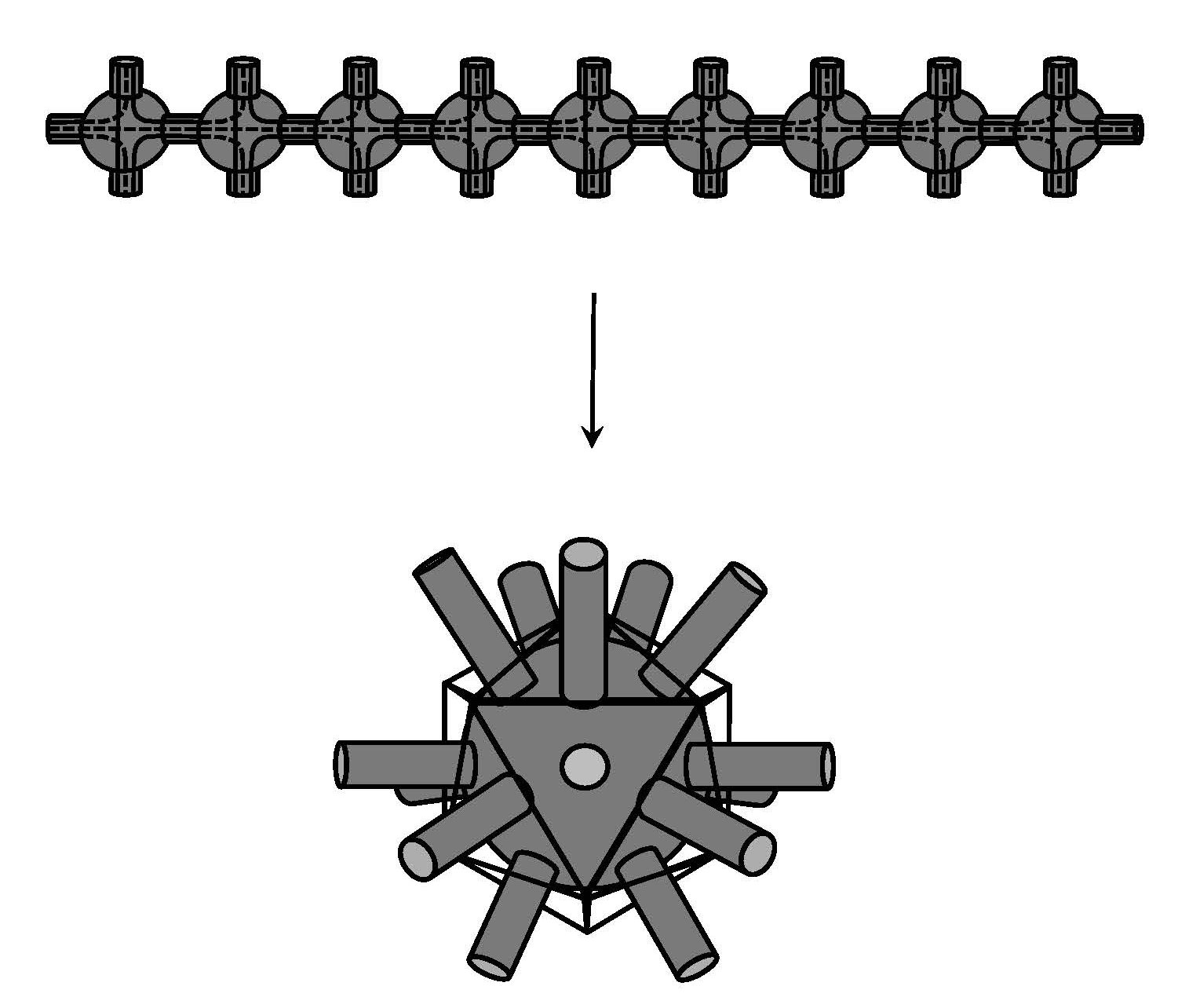}
  \end{center}
\caption{Forming Composite nodes}\label{composite}
\end{figure}

We can then define a $n$ valent BRN with natural valence $m$ (with $n = km-2(k-1), k \in \mathbf{Z}$) as a braided ribbon network where each of the nodes has $n$ tubes which intersect it but where each of the tubes has $m$ racing stripes. Likewise we can define a multi-valent BRN with natural valence $m$ in a similar manner but without fixing the value of $k$ for all nodes. We then construct composite nodes by connecting natural nodes in series by simple edges and shortening the edges which connect them internally until all of these nodes combine into a single sphere with the appropriate number of punctures (see figure \ref{composite}). As these combined nodes are simply glued they are then dual to gluings of simplices which when grouped together would be equivalent to a polygon (for a natural valence of 3) or a polyhedron with triangular faces(for natural valence of 4).

\section{Connecting to Spin Networks}

The states of loop quantum gravity are spin-networks embedded in manifolds up to diffeomorphisms.\cite{Rovelli:2004tv} Though progress has been made studying these objects, much of it has ignored the content of the embedding. We propose that an underlying reason for the lack of study of the embedding of these networks is the difficulty of the problem: each edge of the graph could potentially be linked with any other link, no matter how distant the two are in the discrete distance function on the abstract graph (obtained by counting the smallest number of edges that one must traverse to get between them). On the other hand, in Braided Ribbon Networks we have the reduced link: an operation which extracts from the BRN information regarding the embedding without reference to the individual edges. By connecting spin networks to BRNs we can bring over the reduced link and with it remove the obstacles to studying the content of the embedding of the states of loop quantum gravity.

\begin{figure}[!h]
  \begin{center}
    \includegraphics[scale=0.2]{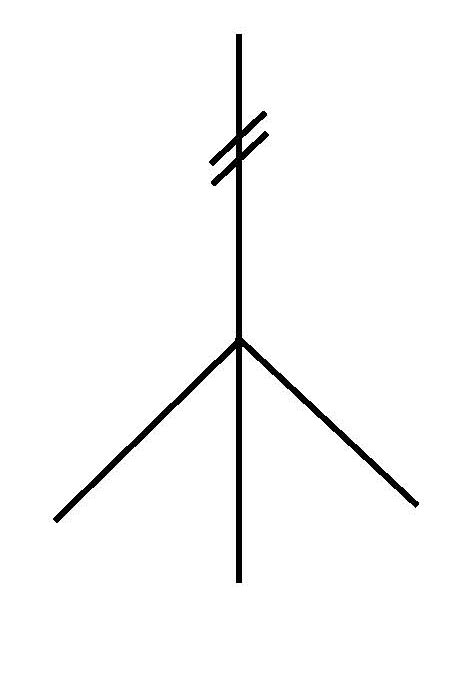}
    \end{center}
\caption{Unambiguous Blackboard Embedding of a 4 Valent Node}\label{bboard}
\end{figure}

\begin{figure}[!h]
  \begin{center}
    \includegraphics[scale=0.2]{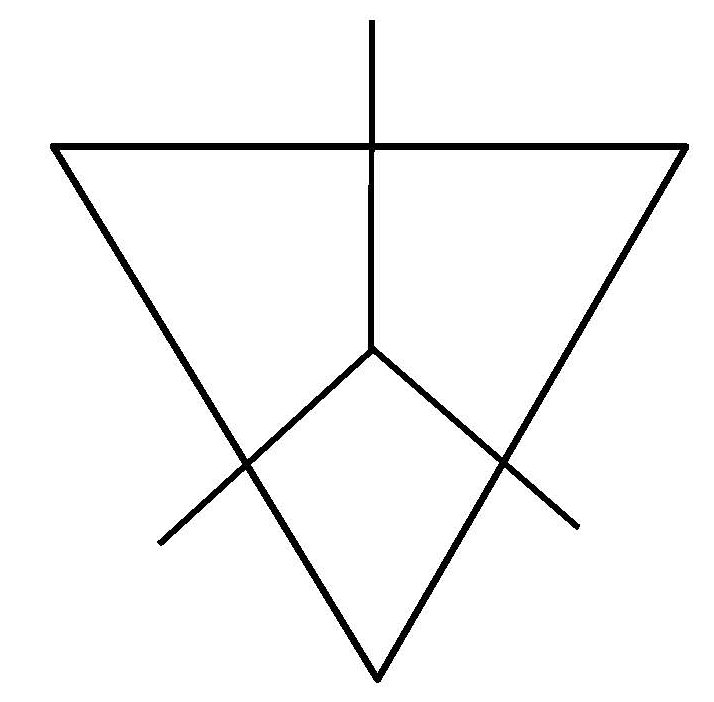}
    \end{center}
\caption{3 Valent Spin Network Node Dual to a Triangle}\label{j1}
\end{figure}

\begin{figure}[!h]
  \begin{center}
    \includegraphics[scale=0.2]{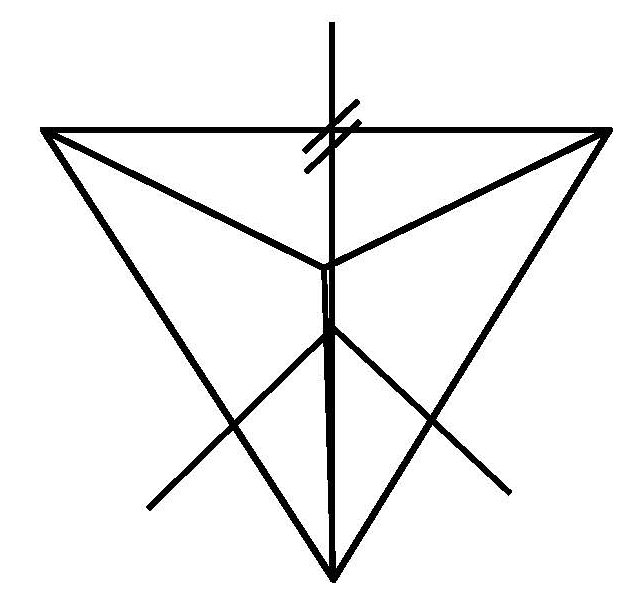}
    \end{center}
\caption{4 Valent Spin Network Node Dual to a Tetrahedron}\label{j2}
\end{figure}

We mention here - for future use - the idea of a blackboard embedding of a graph: an embedding which places the nodes of a graph into a plane and keeps the edges of the graph in the plane except for those locations where they cross one another. Such an embedding is named for the fact that it is how graphs are drawn in practice. Such an embedding introduces ambiguities for a four-valent (or higher) graph, and so we require that a blackboard embedding of such a graph to also include a labeling of each node as shown in figure \ref{bboard} with a slash on one of the incident edges to move it up out of the plane of embedding - giving an orientation to the node. Blackboard embeddings of higher valence graphs can similarly be marked in order to make such an embedding unambiguous. With this ambiguity removed we can define each node in such a graph to be locally dual to a simplex (a triangle for a $3$ valent node, or a tetrahedron for a $4$ valent node) in the same way as we did for BRNs in \cite{Hackett2011}. We do so in the manner prescribed by figures \ref{j1} and \ref{j2}. We also note that instead of this disambiguation we could instead not allow planar embeddings of $4$ valent nodes, but as this paper is constrained to diagrams in two dimensions we will use this notation.

\begin{figure}[!h]
  \begin{center}
    \includegraphics[scale=0.2]{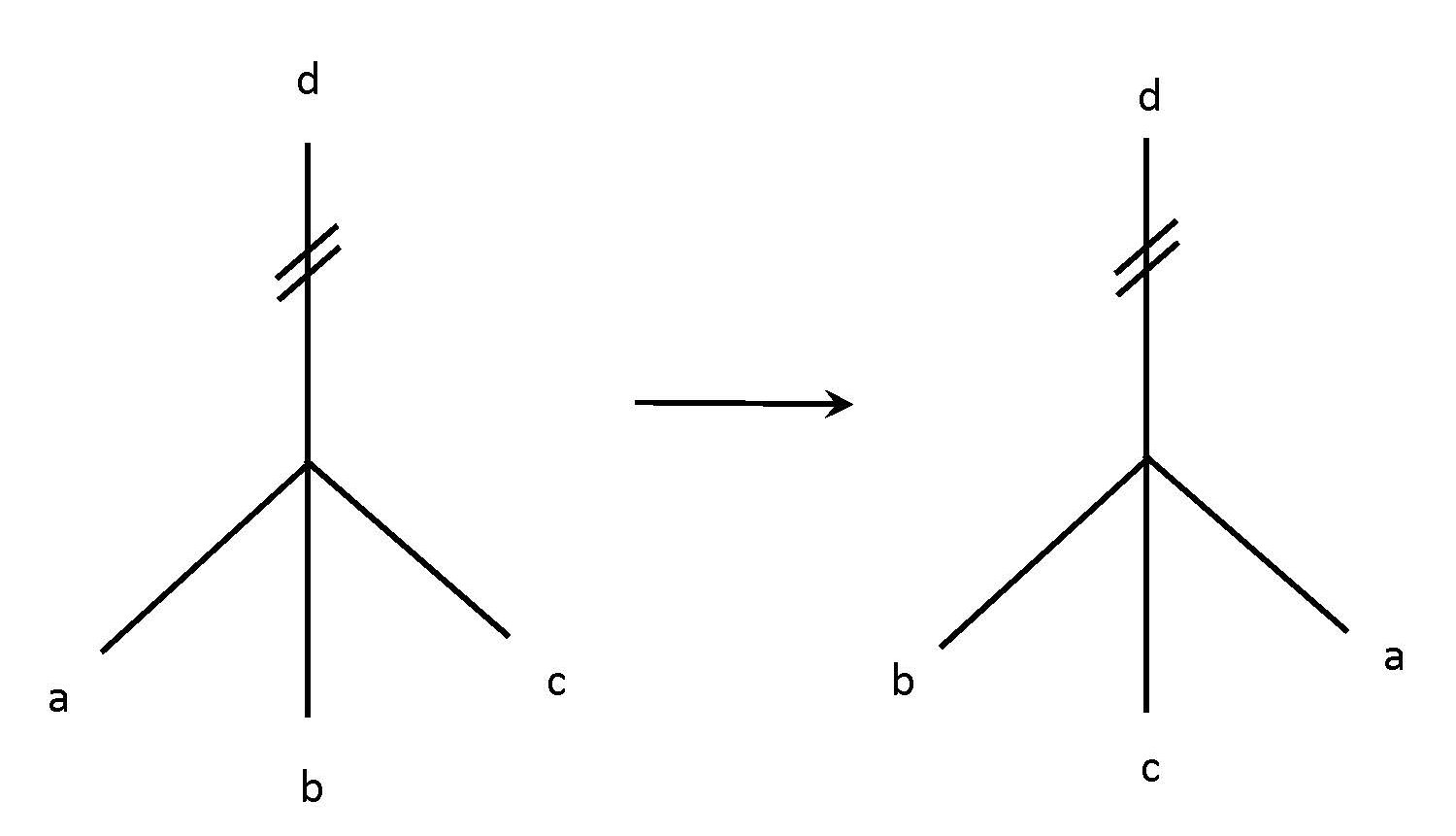}
  \end{center}
\caption{Allowed Ordering Changes}\label{allowed}
\end{figure}

\begin{figure}[!h]
  \begin{center}
    \includegraphics[scale=0.2]{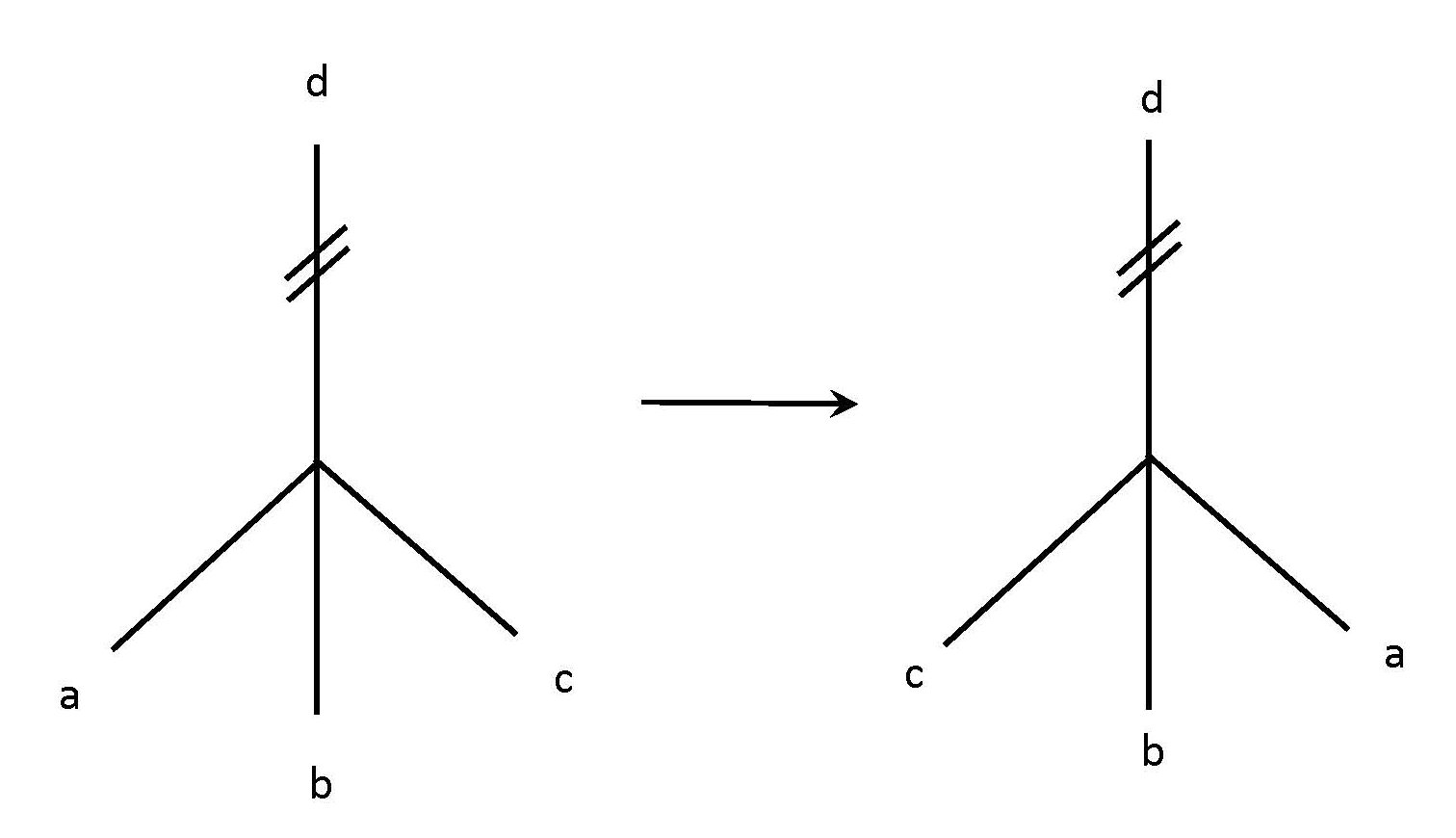}
    \includegraphics[scale=0.2]{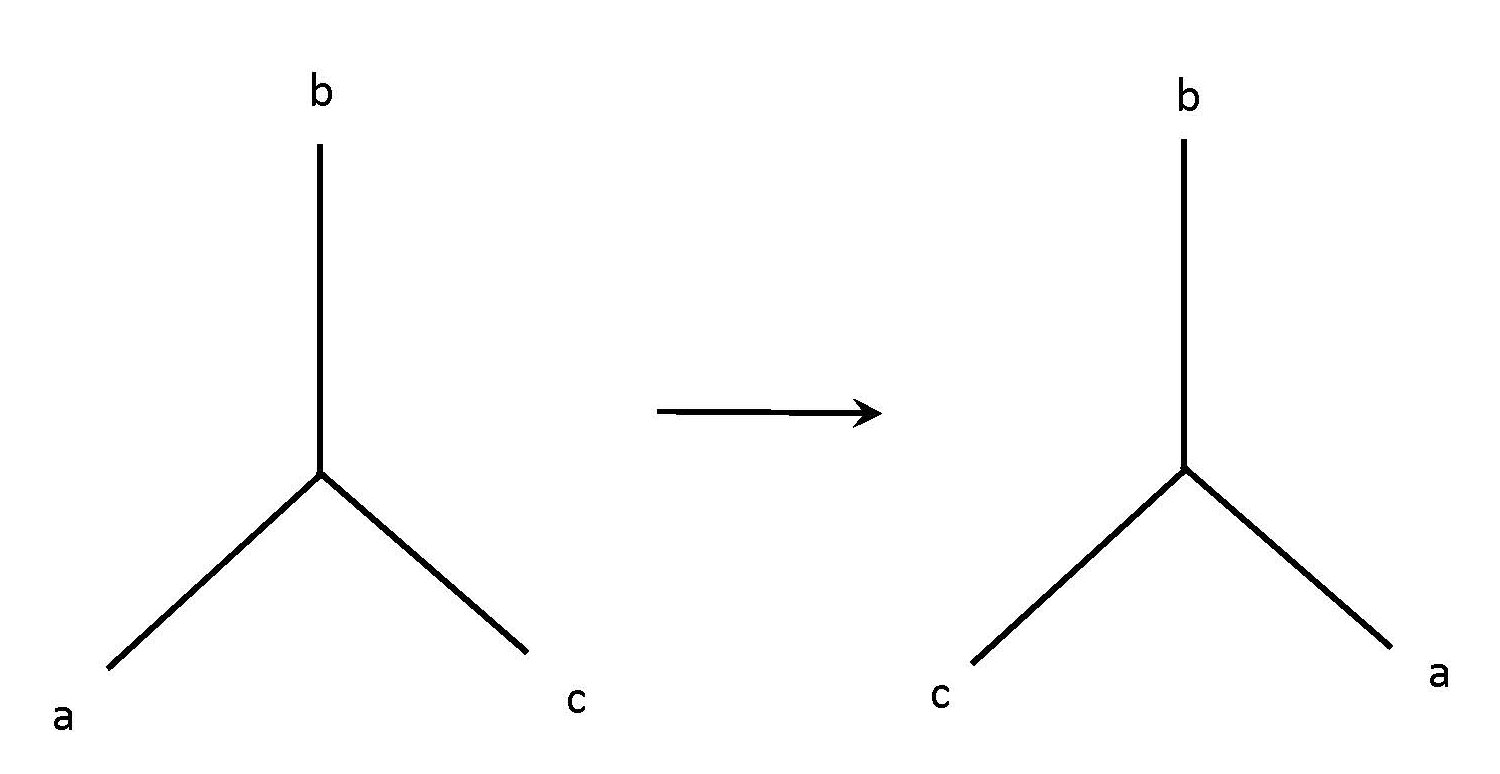}
  \end{center}
\caption{Disallowed Ordering Changes}\label{dis}
\end{figure}

As the content of the states of loop quantum gravity we are concerned with are not dependent on the labelings of the edges or vertices of the graphs, we will ignore these labelings and consider instead only the equivalence classes of graphs of fixed valence embedded in a 3-manifold under the restriction of the diffeomorphism group. The restriction of the group is to omit a class of transformations which change the orientation of the nodes of the graph - that is to say the transformation which takes a graph between two blackboard embeddings which have different cyclic orderings of the edges around each node (for a 3-valent node) or the cyclic orderings of the edges around each node with one edge removed (for a 4-valent node) - these are demonstrated in figure \ref{allowed} and \ref{dis}. We will call such equivalence classes \textit{diffnets} and the different elements of the equivalence class will be called embeddings of a diffnet. We extend this notion to those graphs that don't possess blackboard embeddings by making the requirement that they not change the orientation of the dual simplices of their nodes - this is equivalent to considering the blackboard embedding to be a local property rather than a global one.

\begin{figure}[!h]
  \begin{center}
    \includegraphics[scale=0.2]{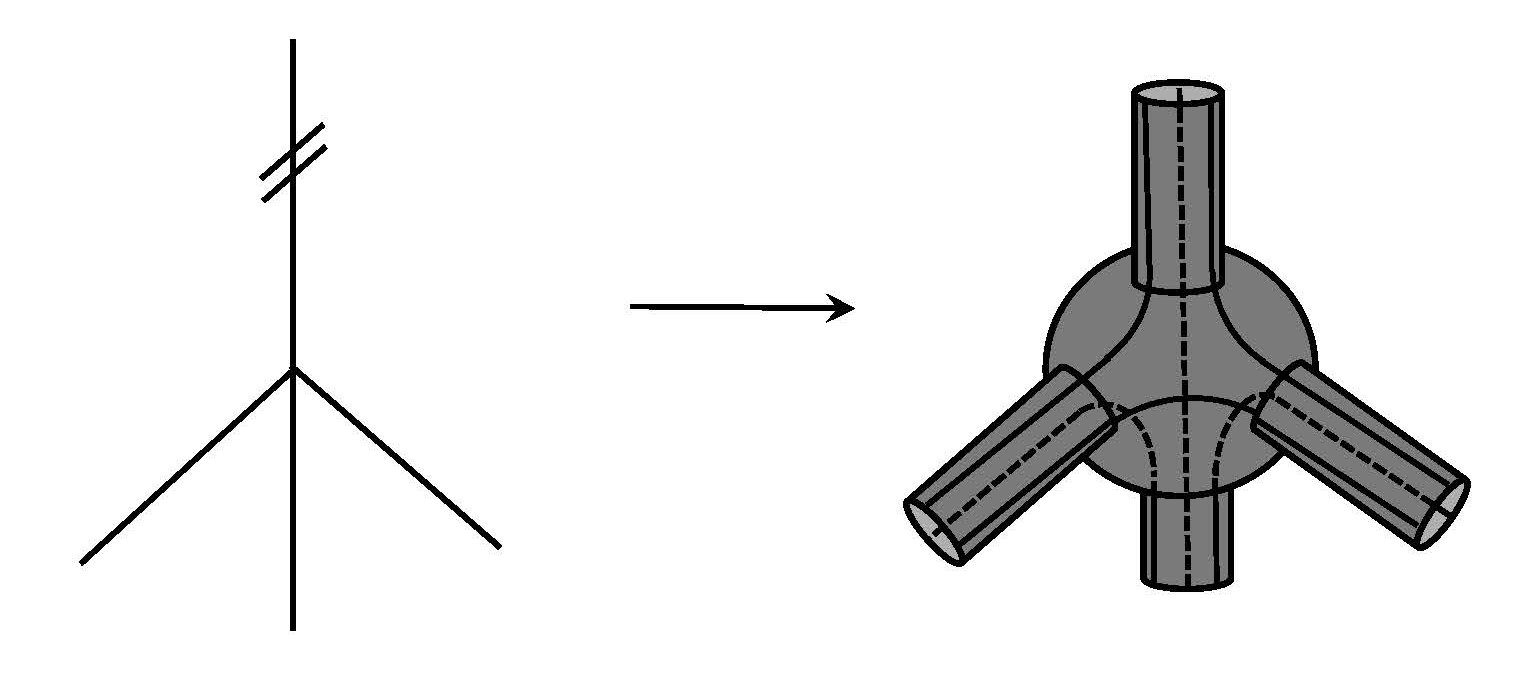}
    \includegraphics[scale=0.2]{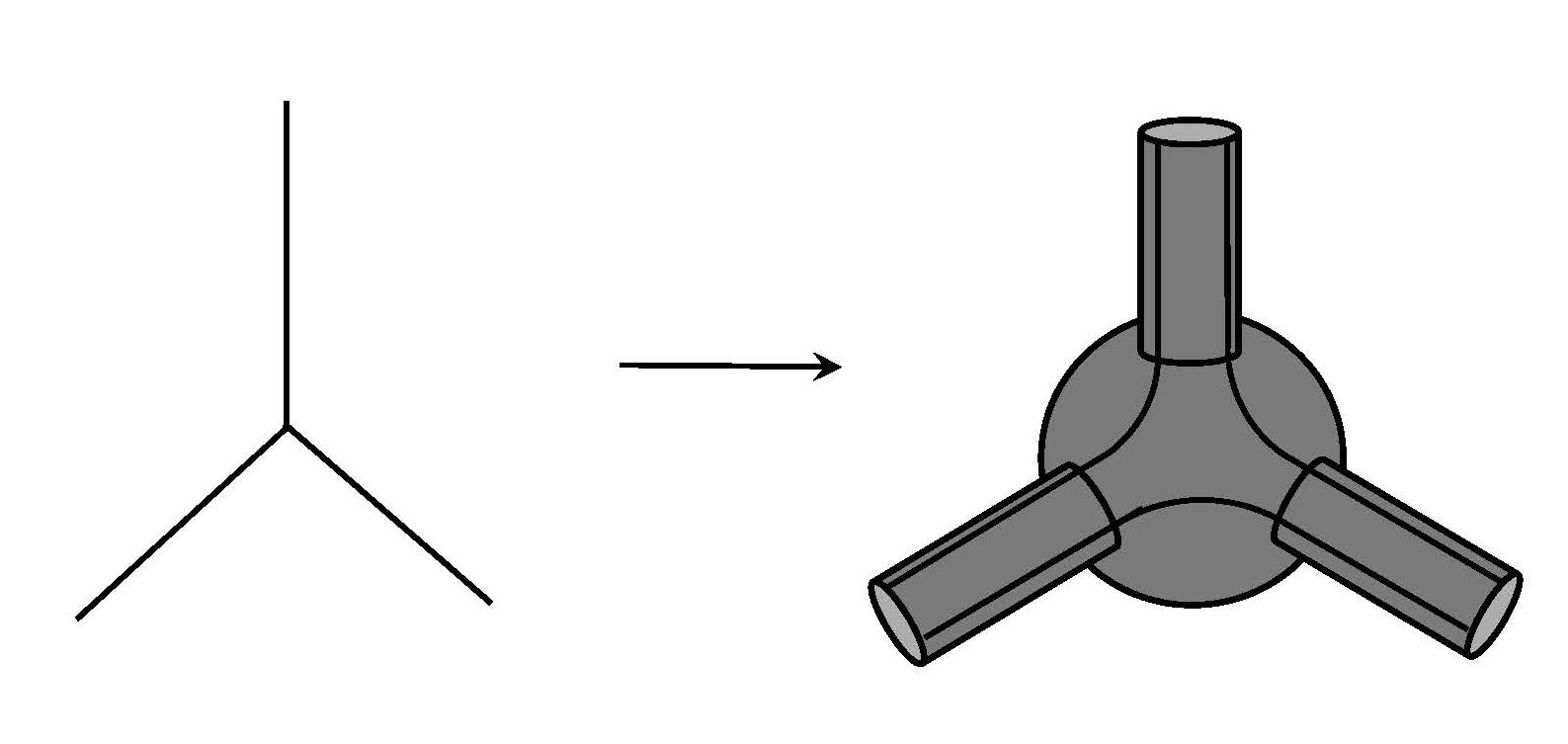}
  \end{center}
\caption{Diffnet to BETBRN Transformation} \label{a1}
\end{figure}

We now set about to demonstrate that there exists a map from $3$ or $4$ valent diffnets in compact manifolds to BRNs. To do so we must first construct a map from a given embedding of a fixed valence graph to an embedding of a BRN. To do this we first observe that since our graph is embedded in a compact manifold that any open cover has a finite subcover. We will construct an open cover as follows: around each node of the graph we take an open ball, around each edge we take an open region which intersects the open balls of the corresponding nodes but does not contain the nodes themselves (we also require that the open region not contain any portion of any other edges), we then take some open cover of the rest of the manifold of which none of the open sets contain any of the nodes or portions of the edges. As the manifold is compact there exists some finite subcover of this open cover, and by construction this subcover will contain each of the balls around nodes and regions around edges. We now define on each part of the open cover (labeled by $i$) a map $F_i$ which will eventually be assembled into a map $F$ on the entire space. On each of the open balls the map $F_i$ takes each node and replaces it with a BRN node which is dual to the same simplex (demonstrated in figure \ref{a1}. For each region containing an edge the map $F_i$ takes the edge and replaces it with a tube with racing stripes. We then require that the choice of racing stripes be one that does not introduce any twisting. This is insufficient to fix the $F_i$ for the edges, however by requiring these $F_i$ to be compatible with those on the open balls around the nodes we can fix them sufficiently. The maps $F_i$ on the open sets which don't contain any portion of the graph are trivial. We thus can construct the map $F$ from this collection of maps $F_i$ as a map from an embedding of a diffnet to an embedding of a BRN.

Our next step is to show that the diffeomorphisms commute with this $F$, by which we mean: that for any diffeomorphism (which preserves node orientation), $\phi$, which takes one embedding of a diffnet to embedding there exists a deformation, $\chi$, between the corresponding elements of the BRN. To demonstrate this we need only realize that any such diffeomorphism can be mapped consistently into a deformation of the corresponding embedding of a BRN. We construct this map as follows: diffeomorphisms are continuous and thus are continuous across the edges (or nodes) of the graph. We can `inflate' this continuity as we `inflate' the edges and nodes - leaving the deformation constant across the surface itself - but giving an identical transformation on the overall structure of the graph. This deformation then exactly mimics the diffeomorphism, barring changing the orientation of a node (which is already forbidden). There is one other form of diffeomorphism which might cause concern - diffeomorphisms which add twists to the edges - and so we will address it specifically.

\begin{figure}[!h]
  \begin{center}
    \includegraphics[scale=0.2]{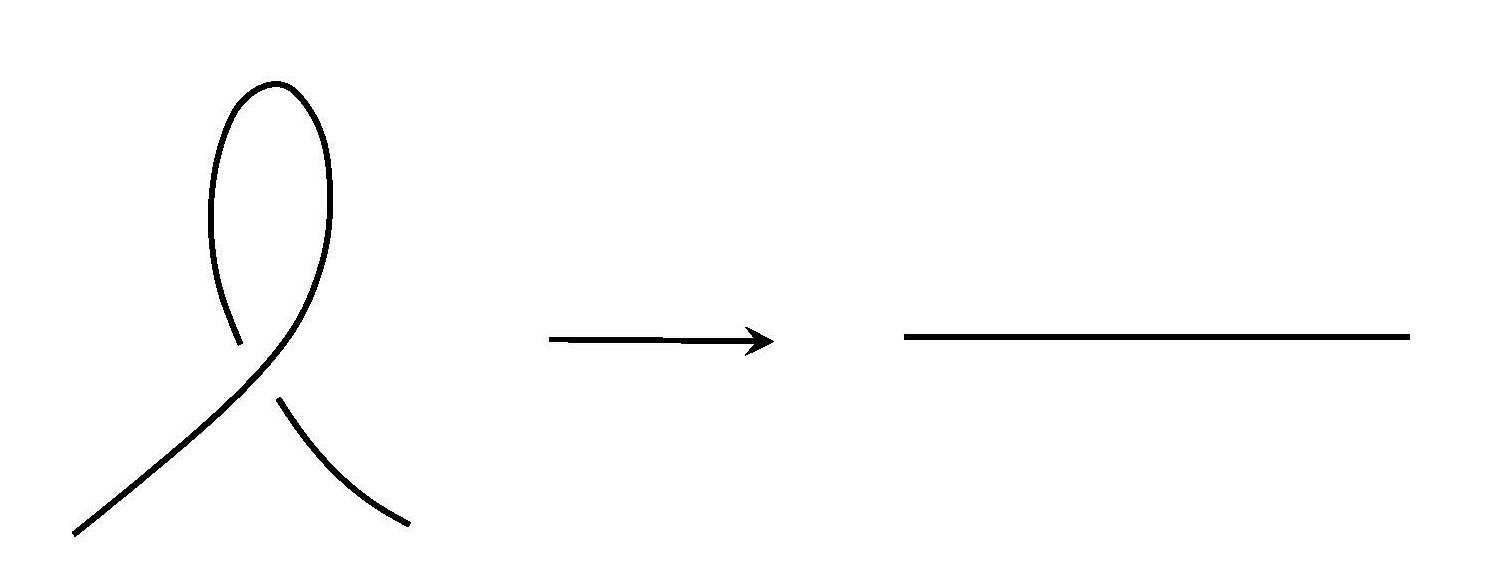}
  \end{center}
\caption{The First Reidemeister Move} \label{a2}
\end{figure}

\begin{figure}[!h]
  \begin{center}
    \includegraphics[scale=0.2]{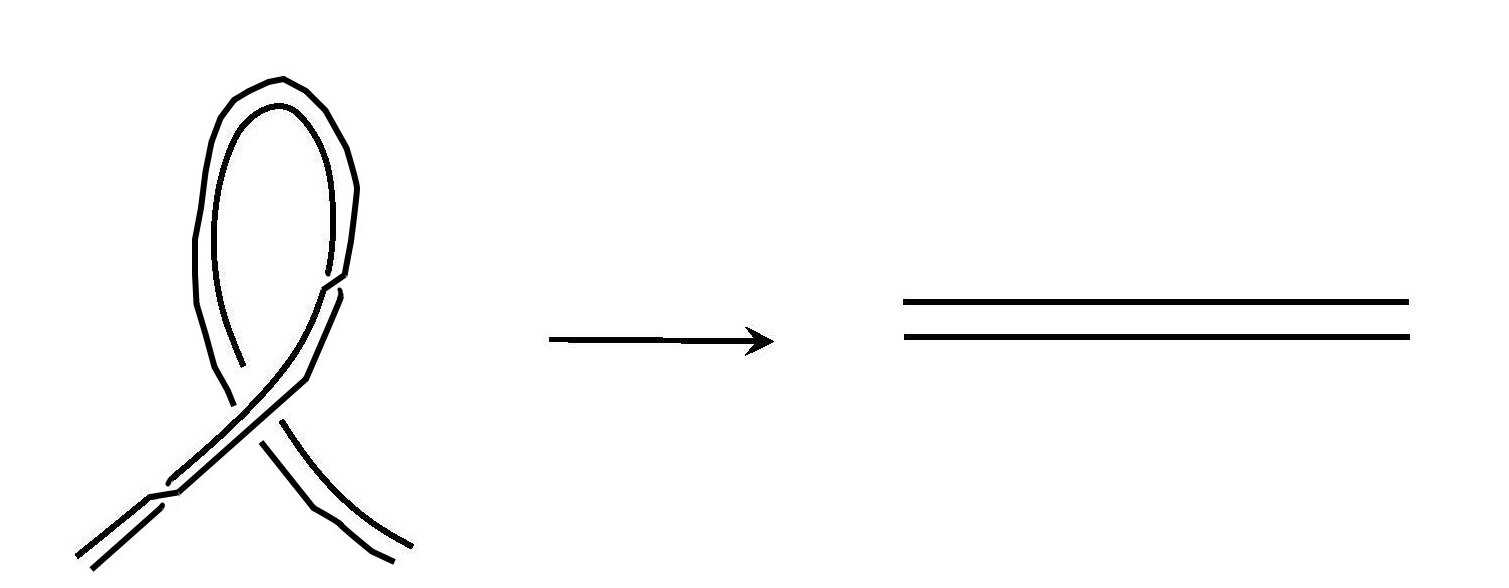}
  \end{center}
\caption{An Untwisted Ribbon} \label{r1}
\end{figure}

\begin{figure}[!h]
  \begin{center}
    \includegraphics[scale=0.2]{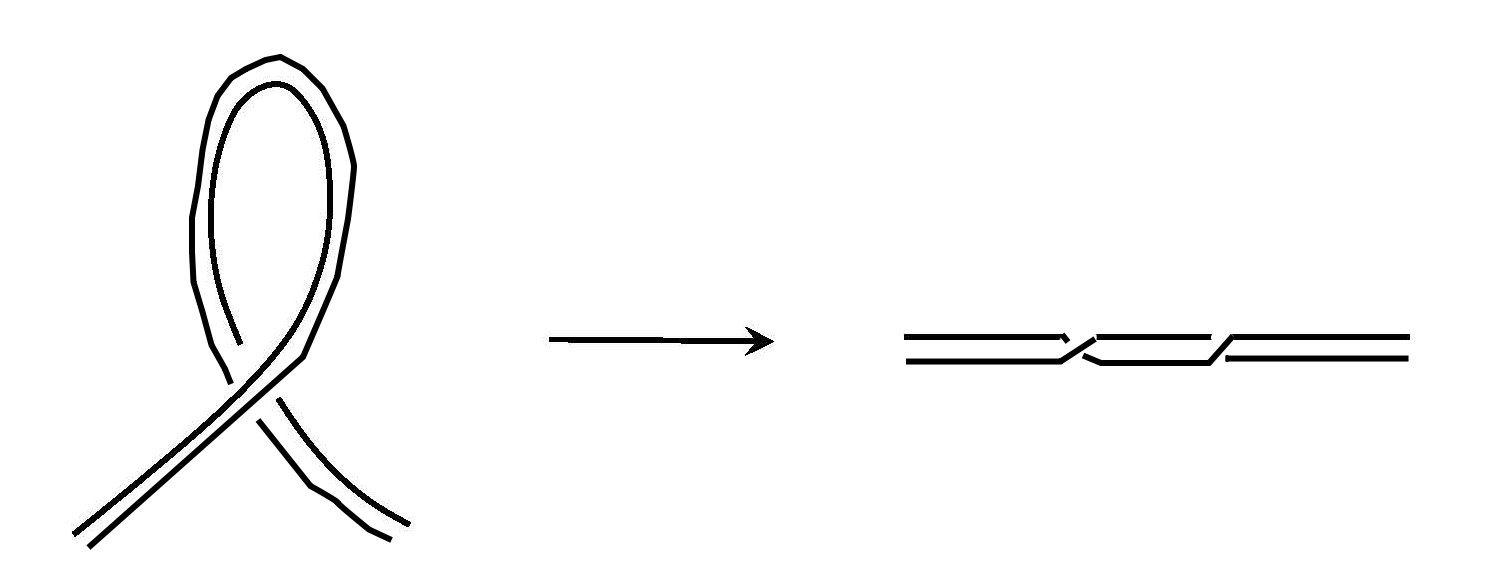}
  \end{center}
\caption{A Twisted Ribbon} \label{r2}
\end{figure}

What we mean by a diffeomorphism introducing twists to edges should be clarified: the edges here are simply curves in space and so cannot have twist in the same way that a tube or any other 3-dimensional object would. Instead here we mean that the diffeomorphism replicates the process described by the first Reidemeister move, shown in figure \ref{a2}, or its inverse. It is tempting to think that it would not be possible to map such a diffeomorphism into a deformation of a BRN or that worse still should we do so we would be left with an operation that introduces twist. This is not the case. As we show in figures \ref{r1} and \ref{r2} (for a pair of racing stripes) the corresponding deformation to this diffeomorphism preserves twist (the same follows in four valent networks with three racing stripes). The source of the confusion here is most likely due to the fact that the ribbon in figure \ref{r1} is the corresponding inflation of the edge in figure \ref{a1}, not the ribbon in figure \ref{r2} as might be people's first instinct.

As $F$ maps every embedding in the equivalence class of a diffnet to an embedding in a single equivalence class of the BRNs, we can promote it to a map on the equivalence classes. That is to say that $F$ maps diffnets to BRNs in the same way that it maps members of these equivalence classes. This gives us an important result: we have that the equivalence classes of spin-networks embedded up to orientation preserving diffeomorphisms can be mapped consistently into the braided ribbon networks.

The point of finding such a relation between diffnets and braided ribbon networks is to bring the machinery developed in \cite{Hackett2011} over to spin-networks. To do this we need to show one further thing: that the evolution moves dual to the pachner moves also commute with the map $F$. This follows immediately if we impose the same requirement on the evolution moves for embedded spin-networks that we used to construct the evolution moves for BRNs: that the orientation of the simplices dual to the nodes is reflected in the orientation of the nodes and that just as that orientation of the external faces is preserved by the pachner moves, it is preserved by the evolution moves.

This allows to state our result more formally: we have shown that there exists a functor from the category of diffnets with morphisms equal to the applications of the evolution moves to the category of BRNs with morphisms equal to the applications of the evolution moves. The category of BRNs can be considered as the sum of its connected subcategories (each corresponding to the set of BRNs which are connected by a series of evolution moves). The reduced link from \cite{Hackett2011} is a functor from the category of BRNs to the category of links embedded in the manifold up to diffeomorphism (with all morphisms being identity morphisms). Due to the lack of non-identity morphisms in this category of links, it assigns the same link to any two connected elements of the BRN category. With these two functors we then construct the composite functor - the reduced link of diffnets.

\section{Conclusion}

We have presented above two results. Firstly, that the embedded spin networks under a restriction of the diffeomorphisms can be consistently mapped to a subset of the braided ribbon networks. Secondly - building from the first result - that the reduced link is an invariant of such classes of spin networks.

These results present a new opportunity in loop quantum gravity: we have the foundations of tools which could be used to study the topological properties of embedded spin-networks. Indeed not only do we have such a tool, but additionally all of the results of braided ribbon networks can be imported to spin networks in general.

The ability to move from spin networks to braided ribbon networks may have other consequences as well. Looking at the origins of braided ribbon networks, it is interesting to note the proposal of \cite{Smolin:2002sz} to use framed graphs for quantum gravity with a positive cosmological constant. It is possible that braided ribbon networks could be the right arena in which to study the relationship of the framed graphs to the standard spin networks.

There still remains work to be done in the context of braided ribbon networks: the degree to which the reduced link captures all of the topological information is still as of yet unknown and a physical interpretation of the majority of the content of the reduced link of BRNs (or likewise, spin networks) is still as of yet unknown. These are the subject of continuing research in the field.

\section*{Acknowledgements}

The author is indebted to his Advisor Lee Smolin, for his
discussion and critical comments. Thanks are also extended to the group at Marseille for hosting the author during the time this work was cemented and for their critical discussion of the work. Additional thanks to Louis Kauffman for discussions and collaboration on related topics. Particular thanks are due to Adam Ramer for his providing the figures contained within. This work was supported by the Government of Canada through NSERC's CGS-MSFSS program and NSERC's CGS program. Research at Perimeter Institute for Theoretical Physics is supported in part by the Government of Canada
through NSERC and by the Province of Ontario through MRI.

\end{document}